A New Signal from the Sun: Maximum Sunspot Magnetic Fields Independent of Solar Cycle

William Livingston, Fraser Watson

National Solar Observatory, 950 N. Cherry Ave., Tucson, AZ 85726, USA.

Abstract:

Over the past five years, 2010-2015, we have observed, in the near infrared (IR), the maximum magnetic field strengths for 4145 sunspot umbrae. Herein we distinguish "field strengths" from "field flux". (Most solar magnetographs measure flux). Maximum field strength in umbrae is co-spatial with the position of umbral minimum brightness (Norton and Gilman, 2004). We measure field strength by the Zeeman splitting of the Fe 15648.5 A spectral line. We show that in the IR no cycle dependence on average maximum field strength (2050 G) has been found ± 20 Gauss. A similar analysis of 17,450 spots observed by the Helioseismic and Magnetic Imager onboard the Solar Dynamics Observatory reveal the same cycle independence ± 0.18 G., or a variance of 0.01%. This is found not to change over the ongoing 2010-2015 minimum to maximum cycle. Conclude the average maximum umbral fields on the Sun is constant with time.

Introduction:
The Sun's waxing and waning surface magnetism causes sunspots and the resulting well known ~11 year cycle in the number of spots. The number of spots, and other cycle indicators, such as 10.7cm radio flux, total irradiance change, etc., are all "magnetic flux" dependent. In this letter we discuss instead "field strength". A sunspot consists of a dark umbra, with larger spots surrounded by a less dark penumbra. All sunspots are the sites of kilo Gauss magnetic fields, with the strongest field strength component at the darkest position in the umbra. This letter deals with the unexpected discovery that, over the past 5 years, the maximum umbral field strength on a given day, averaged over all spots visible on the entire visible disk, is a constant independent on the cycle phase.

Methods and Observations:
Surface magnetic fields are measured by the Zeeman splitting of suitable Fraunhofer spectrum absorption lines. Solar "magnetographs" make use of the fact that the Zeeman components are oppositely polarized and they are able to measure magnetic flux outside of sunspots to very small levels. Our observations reported here do not involve polarization but simply the Zeeman splitting of Fe 15648.5 A. Such splitting is always complete inside spot umbrae, and this is why we can measure actual field strength in Gauss (Fig. 1).

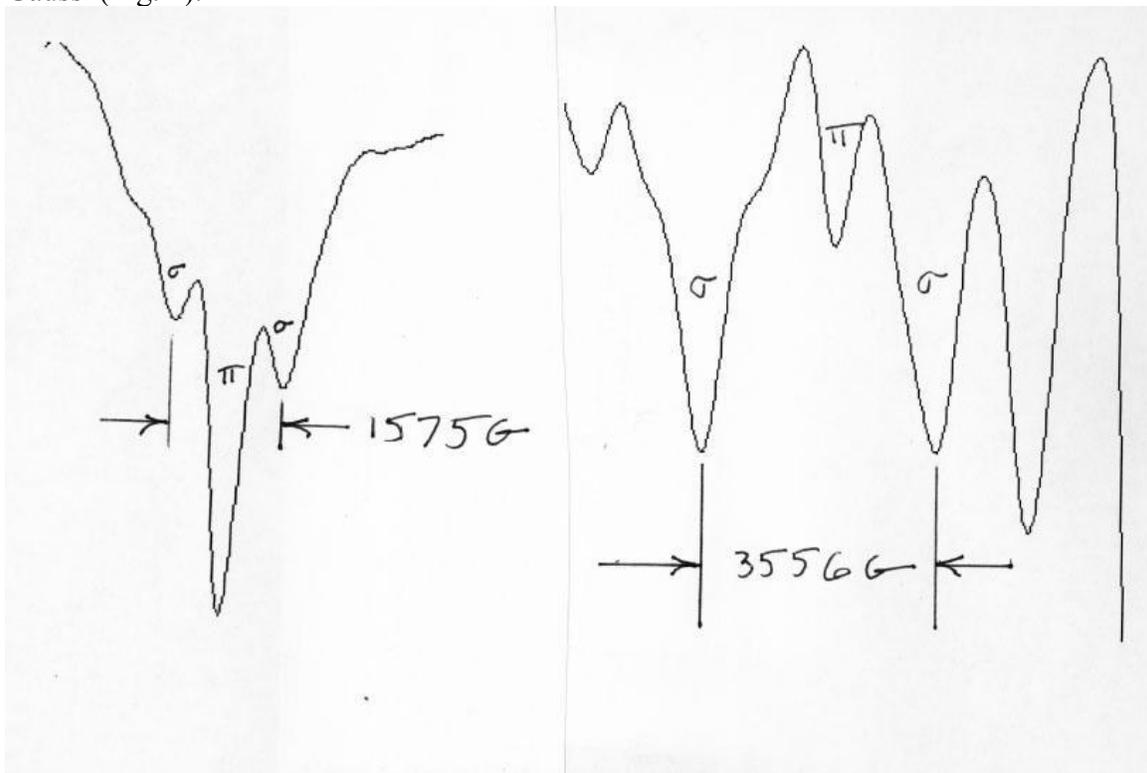

Figure 1. Examples of Fe 15648.5 A Zeeman line splitting. At left is a tiny spot umbra without a penumbra where the field, based on separation of the sigma components, is 1575 Gauss. The so-called pi component for such a small spot is largely scattered light and is ignored here. To the right is the instance of a large spot (with a penumbra) where the sigma separation yields a field of 3556 G. The pi component represents a transverse

field (irrelevant here; as is the strong line to the right which is due to the OH molecule, also irrelevant).

This spectral line is probably the most favorable in the entire electromagnetic spectrum for umbral fields (Solanki, Rüedi , and Livingston, 1992). Its central depth is moderate (leading to a relative line depth of 0.3-.5) in both spot umbrae and the quiet Sun. It has good splitting response (Landé g = 3), and is relatively unblended by other lines. The fact that this is in the near IR is favorable because scattered light is low compared to visible wavelengths. This factor is important for accurate site measurement location (maximum field strength co-spatial with minimum umbral brightness).

Our observations are made on the 0.82 m diameter main image of the McMath-Pierce Solar Telescope in Arizona. First, a pencil drawing is made on which the spots to be observed are identified by National Oceanic Atmospheric Administrastion (NOAA) number and the date (Fig 2).

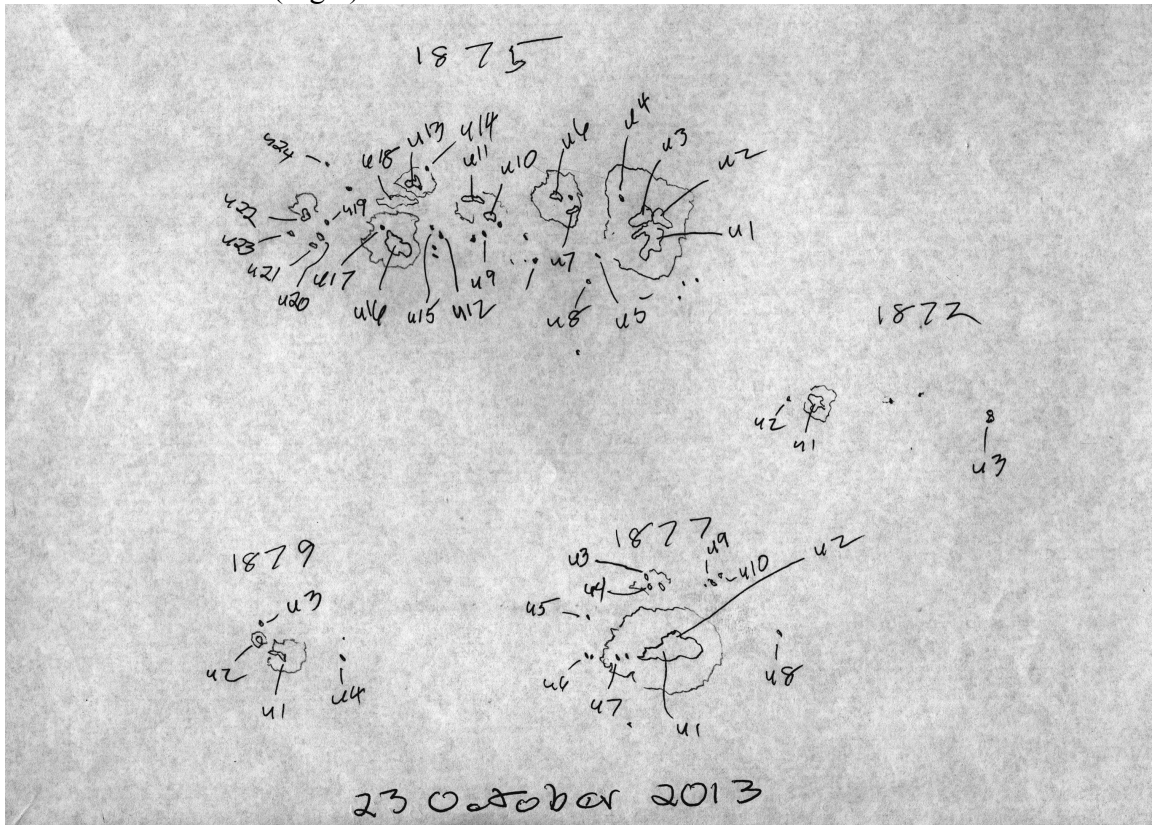

Figure 2. An example of a drawing made at the telescope to identify the exact umbrae observed. The symbol u indicates an umbra actually recorded and in the data file (see acknowledgements). Any spot not given a u number was not successfully observed. The number above each spot group is the NOAA active region number which was given on the internet.

An image slicer of 1 mm square entrance aperture (2.5 x 2.5 arc-sec on the scale of the main solar image) feeds square entrance aperture (2.5 x 2.5 arc-sec on the scale of the main solar image) feeds light to the 13.7 m focal length spectrometer. The detector is a single, liquid nitrogen cooled, InSb diode. As this detector is a single element diode, there is no flat-field requirement as would be necessary for an array detector.

In 1995 we began Fe 15648.5 A spot observations with Solanki, Rüedi, and Livingston (1992) and students (for example Rüedi, Solanki, and Livingston1995). Initially we investigated large spots and the work was confined to those. However, in 2010 we began a more systematic program where we made one measurement at the darkest position of every spot umbra we could see on the solar disk. As mentioned, the darkest umbral position normally yields the strongest field strength in a given spot umbra. Norton and Gilman (2004) found the minimum dark position is the best criterion for maximum field strength. We believe this agrees with sunspot models.

Results:
Results from our observations 2010-2015 are displayed in Fig. 3.

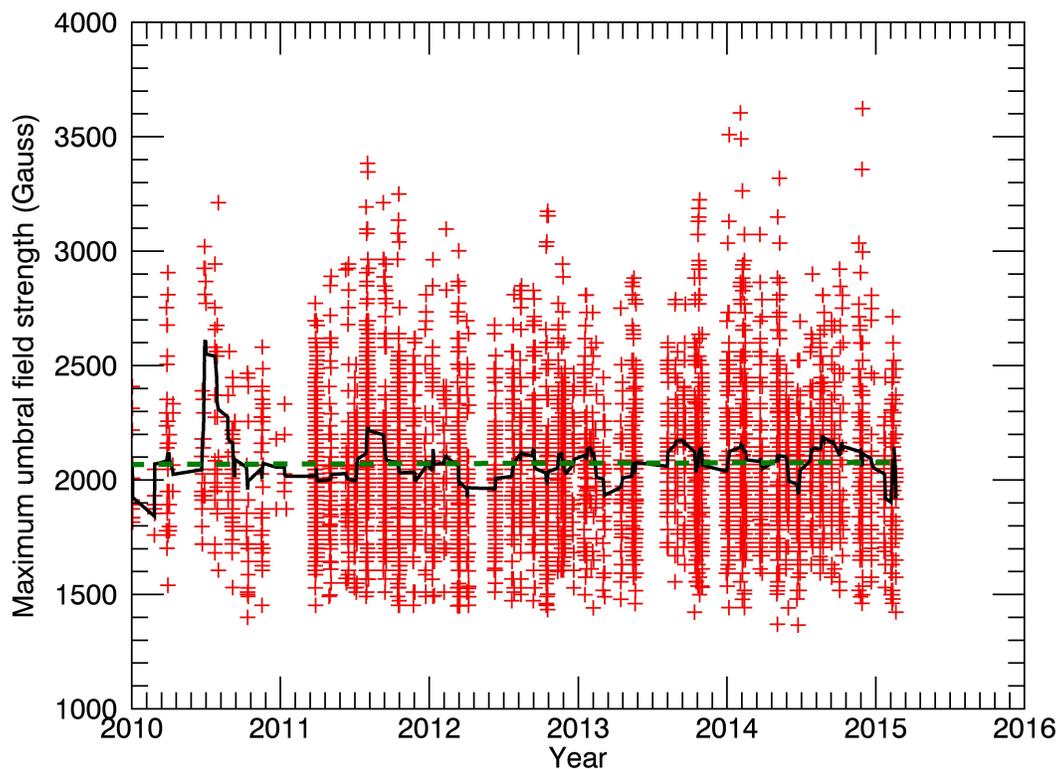

Figure 3. Individual Fe 15648 maximum umbral magnetic field measurements displayed red. A linear fit to this data is flat ± 20.0 G/year and is shown by the dashed green line. The day averages are black (including times of no observation).

Time allocation for this project has been typically 4-5 consecutive days per month. Clear sky near the Sun is essential. Signal/noise is set by image motion, not system or photon noise (which is negligible). In practice the observer searches a spot umbra manually for the darkest position as indicted by an analog signal meter. Each set of 3 scans requires about 4 seconds of time. If the displayed resulting spectrum is satisfactory, and a clear Zeeman splitting signal is seen, it is recorded.

Although there is a large diurnal range in the measured umbral magnetic fields, from around 1500 to 3600 Gauss, the mean umbral magnetic field over time is almost constant (2100 G). This result is surprising as the Sun was very different in 2010, in terms of sunspot numbers, when compared to 2015. In 2010, solar activity was still close to the minimum between solar cycles 23 and 24. In 2014-2015, the Sun has been relatively active with many large sunspot groups as it has been passing through the maximum phase of solar cycle 24. Another representation of the data is by the bi-annular distribution of field strengths as given in Fig. 4.

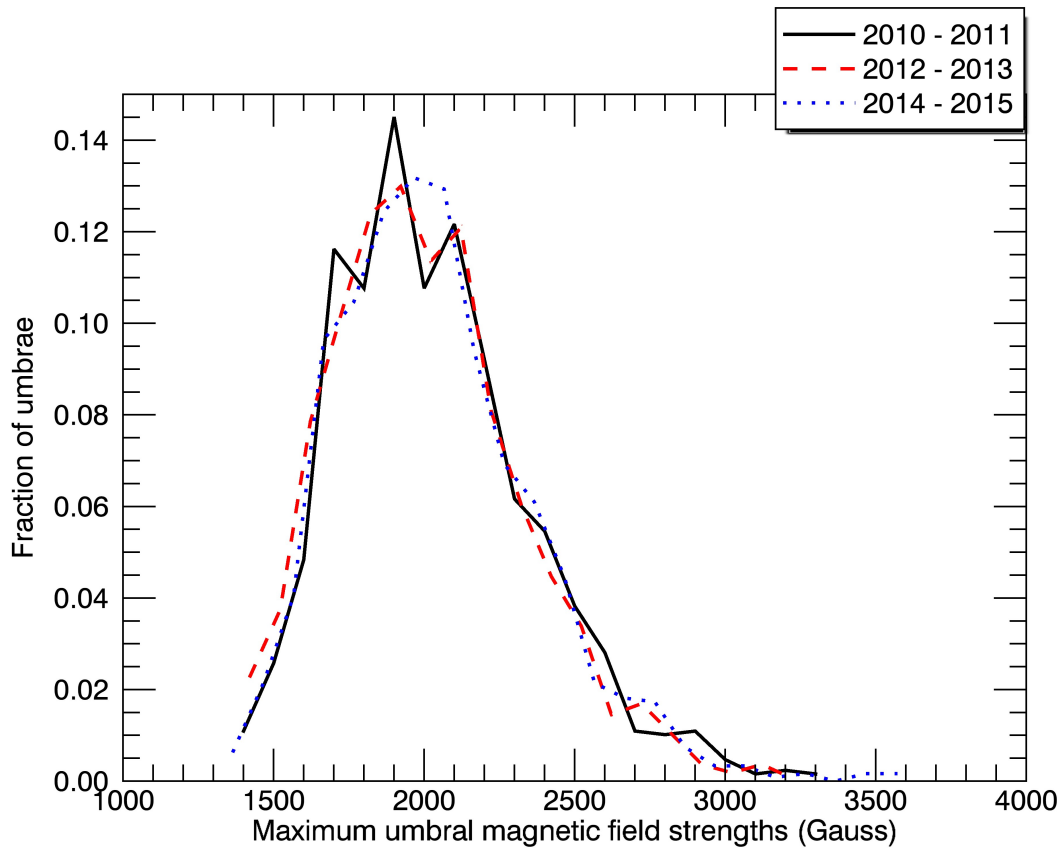

Figure 4. Distribution of field strengths as a function of time. Each distribution is two years wide and normalized such that the area underneath it totals 1.0. The numbers of umbrae in each time interval are comparable (numbers = 1281, 1654, 1241).

Again we note there were fewer spots, and therefore fewer observations, in 2010-2011 than in 2014-2015. Yet the distributions are almost identical. Many of the recent sunspots observed were large with strong magnetic fields. So there must have been many more small, weak sunspots, which seems to offset the stronger sunspots and keep the average umbral magnetic field strength near a constant.

To further confirm the above result, we looked at continuum and magnetogram data from the Helioseismic and Magnetic Imager (HMI) as part of NASA's Solar Dynamics Observatory spacecraft (Schou, J., et al., 2012). This instrument provides a number of advantages over the ground-based measurements: such as no issues with atmospheric seeing, and a far superior daily cadence due to the daily synoptic nature of the HMI measurements. Rather than using a manual detection method, an automated sunspot tracking and detection code known as STARA was employed (Watson, Flecter, and Marshall, 2013). This eliminates some of the possible biases of human observation, such as a change in visual ability of the observer over time. To determine the maximum magnetic field strength of the sunspots detected by this automated algorithm, a Milne-Eddington inversion is used. This technique involves using a model of the solar atmosphere in conjunction with polarimetric sunspot observations to derive the magnetic field at the source (i.e. the darkest umbral position). Fig. 5 shows the maximum umbral magnetic field measurements for the 17450 HMI umbrae.

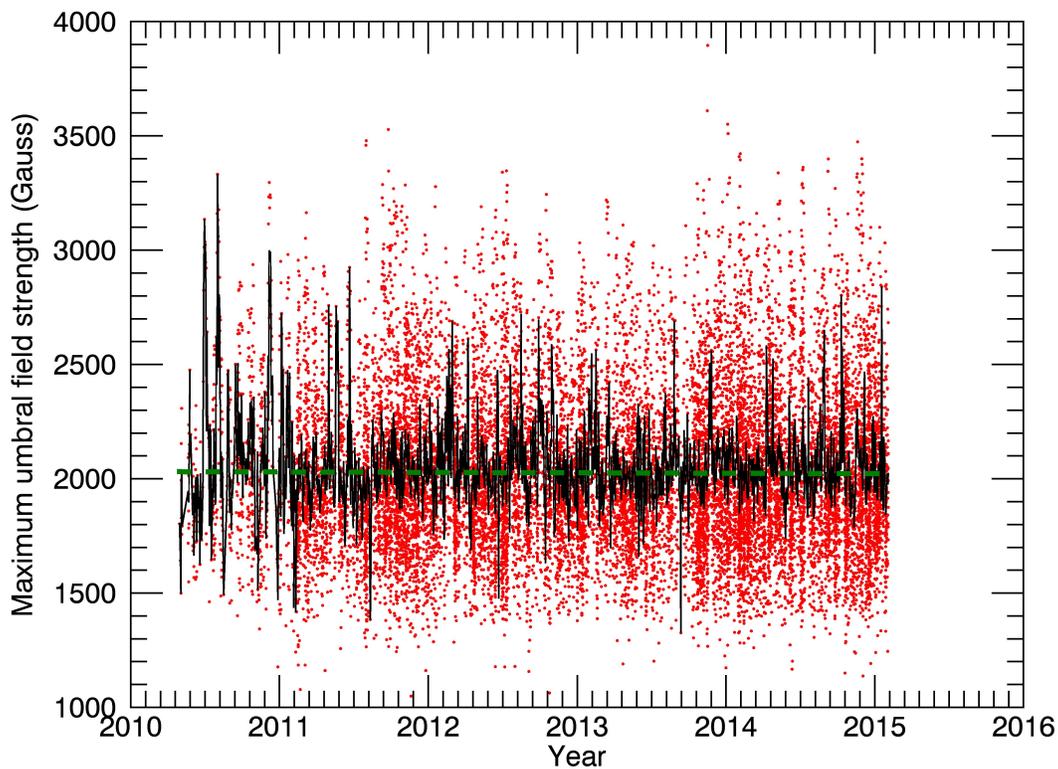

Figure 5. Fe 6173 A individual fields from HMI SDO space telescope; The individual fields are indicated by red dots. Otherwise this Figure is the same as Fig.2.

The result is similar to that was seen in the McMath-Pierce Solar Telescope data for the same time interval. The average field measured remains near constant (2050 G) with a variance of ±0.18 G over time. This means a variance of about 0.01%

The solar activity cycle is driven by emerging magnetic flux which leads to the approximate 11 year variation in sunspot numbers, magnetic flux (Sun-as-a-star), EUV flux, solar irradiance, and so on. One might expect, therefore, to find a cycle modulation in umbral maximum magnetic field strengths, but we do not see any in our observations, Fig. 6.

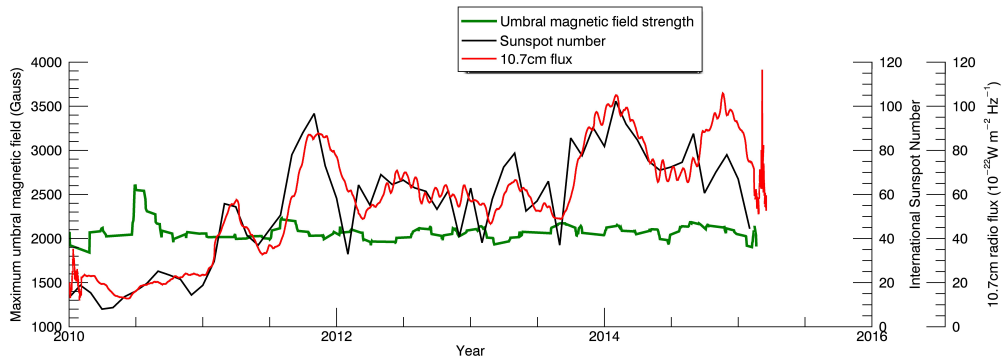

Figure 6. Comparison of field flux data, as indicated by sunspot numbers (black) and 10.7 cm radio flux (red), as compared to IR sunspot maximum field strength data (green) herein presented.

Norton, Jones, and Liu (2013) have pursued a similar quest, also with mostly negative results. We cannot extend our maximum field result back before 2010 because suitable Fe 15648 Å magnetic observations are not available.

Summary:
At high spatial resolution sunspots do differ from each other as shown by brightness structure across the umbra: umbral dots, light bridges, and other details. Our data in this letter has nothing to do with the above structure. We simply measure the magnetic field at the darkest position in each observable spot umbra, see examples in Fig. 1. We repeat this across the entire visible solar disk on a given clear day, as in Fig. 2. We then assign equal weight to each observation without regard to spot size. The results are in Figs. 3-6. Our tentative conclusion is that in the photosphere, where spots are formed, the physical conditions are constant in time. In an earlier study (Livingston, Wallace, White and Giampapa, 2007), covering three cycles 1974-2006, the quiet photosphere near disk center, free of spots, was also found to have constant temperature and independent of the activity cycle. In summary, both the magnetic Sun (as represented by average max sunspot fields) and the quiet Sun outside sunspots are constant with time (2010-2015).